# Prospective performance of graphene HEB for ultrasensitive detection of sub-mm radiation

**Boris S. Karasik [1], Christopher B. McKitterick [2], and Daniel E. Prober [2]**

[1] *Jet Propulsion Laboratory, California Institute of Technology,
4800 Oak Grove Drive, Pasadena, CA 91109, USA
email: boris.s.karasik@jpl.nasa.gov*
[2]*Departments of Physics and Applied Physics, Yale University,
New Haven, CT 06520, USA*

*Noise Equivalent Power and time constant of a submillimeter wave Hot-Electron Bolometer (HEB) made from monolayer graphene are analyzed using the lowest electron-phonon thermal conductance data reported to date. Frequency-domain multiplexed Johnson Noise Thermometry (JNT) is used for the detector readout. Planar microantennas or waveguides can provide efficient coupling of the graphene microdevice to radiation. The results show that the graphene HEB detector can be radiation background limited at very low level corresponding to the photon noise on a space telescope with cryogenically cooled mirror. Beside the high sensitivity, absence of a hard power saturation limit, higher operating temperature, and the ability to read 1000s of elements with a single broadband amplifier will be the advantages of such a detector.*

## 1. INTRODUCTION

As astronomers are planning more powerful instruments for the next generation of submillimeter telescopes, the need for better detectors is becoming more urgent. Several advanced concepts have been pursued in the recent years with the goal to achieve a detector Noise Equivalent Power (*NEP*) of the order of $10^{-20}$-$10^{-19}$ W/Hz$^{1/2}$ that corresponds to the photon noise limited operation of the future space borne submillimeter wave (sub-mm) spectrometers under an optical load $\sim 10^{-19}$ W [1]. Some concepts utilize advanced versions of the bolometer with the Transition-Edge Sensor (TES) thermometer [2], some utilize various forms of quasiparticle detectors read via a change of either the kinetic inductance of a superconducting resonator [3], or

B. S. Karasik, C. B. McKitterick, and D. E. Prober

the capacitance of a small superconducting island [4]. Our recent work has been focusing on the hot-electron TES [5] where a much lower thermal conductance than in a SiN membrane suspended TES could be achieved. This is due to the weak electron-phonon (e-ph) coupling in a micron- or submicron-size hot-electron Ti TES [6,7]. Using this approach, the targeted low *NEP* values have been achieved recently via direct optical measurements [8]. The kinetic inductance detector [9] and quantum capacitance detector [10] demonstrated recently a similar sensitivity as well.

We see nevertheless the possibility to move the state-of-the-art even further. Increase of the operating temperature and the saturation power, and simplification of the array architecture are believed to be important areas of improvement not only for the aforementioned ultrasensitive detectors but also for sub-mm detectors intended for use in photometers and polarimeters where the background is higher (corresponding *NEP* = $10^{-18}$-$10^{-16}$ W/Hz$^{1/2}$). Recently, graphene has emerged as a promising material for hot-electron bolometers [11-13] due to its weak e-ph coupling, extremely small volume of a single-atom thick sensing element and strong Drude absorption of sub-mm radiation. In this paper, we give a detailed analysis of the expected sensitivity and operating conditions in the power detection mode of a hot-electron bolometer made from a few $\mu$m$^2$ monolayer graphene flake which can be embedded into either a sub-mm planar antenna or waveguide circuit via NbN (or NbTiN) superconducting contacts with critical temperature $T_C \approx 10$ K. Compared to the previous analysis [11], most recent data on the strength of the e-ph coupling are used and also the contribution of the readout noise into the *NEP* is explicitly computed. The readout scheme utilizes the JNT allowing for Frequency-Domain Multiplexing (FDM) using resonator coupling of HEBs. The resonator bandwidth and the summing amplifier noise play a defining role in the overall system sensitivity.

## 2. COOLING PATHWAYS IN THE GRAPHENE HEB

In our other work [13,14], we analyzed the single-photon operation of a graphene HEB. The photon counting mode with bolometers is possible when the photon arrival rate $N_{ph}$ is low compared to the device speed $\tau^{-1}$ and the electron energy fluctuation $(C_e k_B T_e^2)^{1/2}$ ($T_e$ is the electron temperature) in the device is small compared to the photon energy, $h\nu$ ($C_e$ is the electron heat capacity). This mode provides better signal-to-noise-ratio than the power detection due to the possibility of amplitude thresholding of the detector noise [15]. For large values of $N_{ph}$ or/and small photon energies, the power detection mode is used. In this paper we focus on the power detection only. The transitional

**Prospective performance of graphene HEB**

case between the photon counting and power detection still needs to be studied.

For sensitive HEB sensors, three possible paths for electron cooling are usually considered: electron-phonon relaxation [16], electron diffusion [17], and microwave photon emission [18]. In the TES HEB, the e-ph thermal conductance and the associated Thermal Energy Fluctuation (TEF or "phonon") noise usually dominate [7]. The electron diffusion can be practically eliminating by fabricating a sensor with superconducting contacts made from a material whose energy gap $\Delta >> k_B T$. The microwave photon emission conductance $G_\gamma \approx k_B B$ ($B$ is the effective bandwidth of a low-pass filter between the detector and a cold absorber for $f << k_B T/h$) can be engineered to be small, as the signal bandwidth in sensitive detector rarely reaches 100 kHz. In graphene, the situation is quite different however. Because of the absence of the resistance temperature dependence the most practical readout scheme is to monitor the change of the electron temperature via a change of the Johnson noise [11,19-21]. We assume that the HEB is coupled to a broadband low noise GHz amplifier (e.g., [22]) via a resonating transformer with the bandwidth $B$. This leads to non-negligible values of $G_\gamma$.

The hot-electron effect in graphene is well justified at sub-Kelvin temperatures. As in many metal films, the strong electron-electron interaction [23] leads to fermization of the electron distribution function thus allowing for introduction of the electron temperature. The thermal boundary resistance is, in turn, very low compared to the e-ph thermal resistance [12]. This allows for consideration of the thermal dynamic in graphene using a simple thermal model considering only cooling of electron subsystem to the phonon bath with constant temperature $T$. The coupling between electrons and acoustic phonons in graphene has been studied theoretically and experimentally in recent years. The summary of experimental data on the e-ph thermal conductance $G_{e-ph}$ is presented in Table I. As one can tell, the scattering of values significant and the temperature dependence $G_{e-ph}(T) \sim T^p$ varies, $p$ = 2-3.5. Definitely, more work is needed to understand the effects of doping and fabrication techniques. Nevertheless, we attempt to make an estimate using the lowest $G_{e-ph}$ data [19]. These data were obtained using Chemical Vapor Deposition (CVD) grown graphene, whereas the rest of data were obtained on pristine (exfoliated) graphene. The CVD technique is the most promising as it already yields commercial size wafers (e.g., see www.graphenea.com). In the following, we will use an expression $G_{e-ph}(T) = 4\Sigma A T^3$ [11,12,19] assuming the device area $A$ = 5 $\mu m^2$ and $\Sigma$ = 0.5 mW/(m$^2$ K$^4$) [19]. This uniquely low $\Sigma$ value sets the very low $NEP$ value not found in other bolometric detectors.

### 3. EXPECTED SENSITIVITY AND TIME CONSTANT

Now we derive an expression for the $NEP$. Usually, TEF noise and Johnson noise are considered as fundamental sources of noise in bolometers [24]. The TEF noise spectrum expands up to the cutoff $\sim \tau^{-1}$ above which the Johnson noise remains (see, e.g., [25,26] for superconducting HEB). Since the JNT based FDM should utilize the GHz spectral range (more channels per single summation amplifier) only Johnson noise and the amplifier noise will contribute to the readout noise. The spectral density of this noise is proportional to $(T_e+T_a)$, $T_a \approx 0.5$ K is the amplifier noise temperature [22,29,30]; the theoretical limit for this quantity is $\sim hf/k_B$. According to the Dicke formula [31] in the limit $f < k_BT/h$, the spectral density of the temperature fluctuations in the post-detection bandwidth $\Delta f = 1/(2t_{av.})$, where $t_{av.}$ is the averaging time, is given as follows:

$$\sqrt{\langle(\Delta T_{JNT})^2\rangle_f} = \frac{T_e+T_a}{\sqrt{Bt_{av.}}} \Big/ \sqrt{\Delta f} = \frac{T_e+T_a}{\sqrt{2B}}. \quad (1)$$

Table I. Electron-phonon thermal conductance

| $T$ (K) | $G_{e-ph}/A$ (mW K$^{-1}$ m$^{-2}$) | | | | |
|---|---|---|---|---|---|
| | [11] | [12] | [19] | [27] | [28] |
| 0.1 | 0.07 | 0.0067 | 0.0005-0.002 | 0.6 | 0.05 |
| 1 | 70 | 19 | 0.5-2 | 60 | 50 |

For high-frequency readout, $f > k_BT/h$ ($f > 2$ GHz for $T_e = 0.1$ K) the contribution of the Johnson noise will be less than that given by Eq. 1.

Using the temperature responsivity for the small signal regime ($T_e \approx T$): $S_T = dT_e/dP = G^{-1} = (G_{e-ph}+G_\gamma)^{-1}$ we obtain the following expression for $NEP_{JNT}$ associated with the JNT readout process:

$$NEP_{JNT} = (T+T_a)\big[G_{e-ph}(T)+G_\gamma\big]\Big/\sqrt{2B}. \quad (2)$$

The electron temperature fluctuates intrinsically so the TEF noise is added in the post-detection bandwidth $\Delta f$. In equilibrium ($T_e \approx T$), for $f < \tau^{-1}$, its squared spectral density is given by [26]:

$$\langle T_e^2\rangle_f + \langle T^2\rangle_f \approx 4k_BT^2/G \quad (3)$$

From Eq 3 and the expression for $S_T$ we obtain a well-known equation for the intrinsic $NEP_{TEF}$ present in any bolometer: $NEP_{TEF} = (4k_BT^2G)^{1/2}$.

The thermal time constant is determined by the total thermal conductance: $\tau = C_e/G$. We calculate the electron heat capacity as

## Prospective performance of graphene HEB

$C_e = \left(2\pi^{3/2} k_B^2 n^{1/2} T A\right)/\left(3\hbar v_F\right) = 5\times10^{-21} T$ J/K [11] ($v_F$ is the Fermi velocity) assuming the electron density $n = 10^{12}$ cm$^{-2}$ as in [19].

$NEP_{TEF}$ and $NEP_{JNT}$ are plotted in Fig. 1 as functions of $T$ for two values of $B$ (100 kHz and 1 MHz). The effect of the microwave photon cooling is present only below 0.2 K. It is also reflected in the weakening of the time constant's temperature dependence. Overall, the $NEP$ due to the readout dominates for the given detector parameters. Still very low $NEP$ values can be achieved at relatively high temperature, e.g., $NEP \sim 10^{-19}$ W/Hz$^{1/2}$ at 300 mK and $NEP \sim 10^{-17}$ W/Hz$^{1/2}$ at 1 K The effect of bandwidth $B$ on the readout noise is more significant: larger bandwidth reduces $NEP_{JNT}$ almost everywhere above 0.1 K.

Now we consider the operation of the detector with an optical load $P$. From the expression for the photon-noise limited $NEP_{ph} = \sqrt{2Ph\nu}$ one can see that the background limited operation for $\nu$ = 1 THz with $NEP \approx 10^{-20}$ W/Hz$^{1/2}$ can be already achieved at $P \approx 10^{-19}$ W ($T$ = 0.1 K, $B$ = 1 MHz). At higher power levels, the electron temperature gradually increases as follows from the heat balance equation: $T_e = \left(T^4 + P/\Sigma A\right)^{1/4}$. However, this does not lead to hard saturation as in, e.g., TES bolometers. The detector power response becomes non-linear but monotonic and well defined. An increase of the Johnson noise power at the amplifier input, $Nk_B T_e B$, ($N$ is the number of pixels simultaneously read by a summing amplifier) can be an issue for some amplifiers with low dynamic range (SQUIDs, Josephson parametric amplifiers). Here a kinetic inductance parametric amplifier [22] is a suitable readout with a large bandwidth $\approx$ 6 GHz and a large saturation power of -52 dBm (6.3 $\mu$W). With this amplifier, 1000s of graphene HEB pixels can be potentially read.

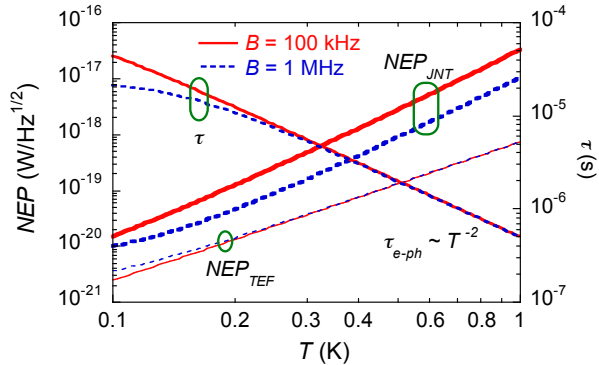

Fig. 1. $NEP_{TEF}$, $NEP_{JNT,}$ and time constant as functions of operating temperature and readout bandwidth.

In conclusion, the hot-electron bolometer based on a monolayer graphene can be a very sensitive sub-mm radiation detector. Even though the readout noise dominates, very low *NEP* values can be expected at higher temperature than those where current sensitive detector operate. Operation of the detector under optical load is not limited by hard saturation effects.

The work at the Jet Propulsion Laboratory, California Institute of Technology, was carried out under a contract with the National Aeronautics and Space Administration. The work at Yale was supported by NSF Grant DMR-0907082, an IBM Faculty Grant, and by Yale University.